\documentclass[12pt,pra,notitlepage,amssymb,tightenlines,showpacs]{revtex4-1}

\usepackage{graphicx}
\usepackage{amsmath}
\usepackage{verbatim}
\usepackage{subfigure}
\usepackage{color}

\begin{document}

\title{Implementation of controllable universal unital optical channels}

\author{A. Shaham}
\affiliation{Racah Institute of Physics, Hebrew University of
Jerusalem, Jerusalem 91904, Israel}
\author{T. Karni}
\affiliation{Racah Institute of Physics, Hebrew University of
Jerusalem, Jerusalem 91904, Israel}
\author{H. S. Eisenberg}
\affiliation{Racah Institute of Physics, Hebrew University of
Jerusalem, Jerusalem 91904, Israel}

\begin{abstract}
We show that a configuration of four birefringent crystals and
wave-plates can emulate almost any arbitrary unital channel for
polarization qubits encoded in single photons, where the channel
settings are controlled by the wave-plate angles. The scheme is
applied to a single spatial mode and its operation is independent of
the wavelength and the fine temporal properties of the input light.
We implemented the scheme and demonstrated its operation by applying
a dephasing environment to classical and quantum single-photon
states with different temporal properties. The applied process was
characterized by a quantum process tomography procedure, and a high
fidelity to the theory was observed.
\end{abstract}


\maketitle
\section{Introduction}

The wave-like nature of a quantum system is manifested by the
existence of a well defined phase between its components. When the
system is isolated, this phase is well defined. However, when the
system interacts with its surrounding, the uncertainty of the phase
may increase in a process called 'dephasing' \cite{Nielsen}. When
quantum information is encoded in the phase, dephasing processes are
interpreted as an addition of noise to the stored information. As a
result, quantum information protocols that rely on the certainty of
the phase may slow down, and their success may be hindered (see for
example \cite{Cirac_1999}).

Dephasing processes belong to the family of unital channels, which
is a special class of decohering processes. A general decohering
processes reduces the information that is stored in a quantum state,
where a unital channel also preserves the average of the input
state. Thus, unital channels are connected with interactions that do
not include an energy dissipation from the state to its
surroundings. Dephasing channels are perhaps the most common type of
the unital channel class since they occur naturally in atomic and
solid state quantum systems (where they are commonly characterized
using the $T_2$ decoherence time scale). In general, photonic
implementations of quantum systems are more immune to decohering
processes, including dephasing, since they only weakly interact with
the environment. Thus, they serve as an appealing candidate in a
variety of quantum information schemes \cite{Obrien_2009}. The noise
robustness of single photons also makes them suitable for studying
decohering processes: transmitting photons through a channel that
successfully induces a certain type of decohering process will not
be accompanied by other types of decoherence to the photons. Hence,
photons are suitable for the demonstration of the effects of
different types of decohering processes on quantum systems and
protocols.

In this work, we study a realization of unital channels that are
applied to quantum bits (qubits) encoded in the polarization of
single photons. Previously, different types of unital channels were
implemented in several ways. Some works investigated the
implementation of a general unital channel while other works
implemented a specific channel type of the unital class, which
mainly was a dephasing channel: general unital channels which were
applied on single photons where implemented using scattering from
different elements \cite{Puentes_2005}. Scrambling schemes that used
Pockels cells \cite{Ricci_2004}, liquid crystals \cite{Chiuri_2011},
or mechanical stress on optical fibers \cite{Karpinski_2008} were
also used to construct different types of unital channels.
Concentrating in dephasing schemes, a channel that completely erases
the polarization phase of the photons was implemented using a single
birefringent crystal \cite{Kwiat_Dephasing}. Control over the
dephasing level of such a channel was achieved by changing the
birefringent crystal length \cite{Xu_2009,Liu_2011}. Dephasing
processes could also be emulated using scrambling schemes that use
wave-plates \cite{Amselem_2009b} or liquid crystals
\cite{Adamson_2007a}. Another method to implement a controllable
dephasing channel is to couple between the polarization and two
different spatial modes via a Sagnac interferometer
\cite{Almeida_ESD} or a polarizing tunable beam displacer
\cite{Urrego_2017}. Here, we study a controllable scheme that is
composed of four birefringent crystals and wave-plates
\cite{Shaham_iso_depo}. Previously, this scheme was only used to
implement an isotropic depolarizing channel. Using a numeric search
we now show that the four crystal scheme can emulate almost every
arbitrary unital channel. We demonstrate the scheme operation by
applying it as a dephasing channel on classical and quantum
single-photon wave-packets. The measured processes are characterized
using a quantum process tomography (QPT) procedure
\cite{Chuang_1997} and present a high fidelity to dephasing
processes. Unlike dephasing schemes that are composed of one or more
birefringent crystals in a fixed orientation, the four crystal
configuration has the advantage that its dephasing magnitude is
known in advance and is not affected by the details of the temporal
structure, such as the coherence time of the initial photonic
wave-packet. This property was verified by applying the dephasing
scheme to two photonic states that differ in their temporal
properties. As expected, the two photonic states experienced the
same dephasing magnitude and agreed well with the theoretical
prediction.

The structure of this article is as follows: in Sec. 2 we give a
theoretical background on dephasing channels and their
representations. The experimental setup for the generation and
detection of single-photon states is presented in Sec. 3. Section 4
is dedicated to a theoretical study of the four birefringent crystal
scheme, and for the demonstration of its operation as a dephasing
channel for two photonic qubit wave-packets that differ in their
coherence time. We summarize the results in Sec. 5. A complementary
study of the temporal differences between the two photonic
wave-packets that was performed using a Soleil-Babinet Compensator
(SBC) is presented in the appendix.

\section{Theoretical background}

The state of a polarization qubit can be described either by the
density matrix $\hat{\rho}$, or by a point in the Poincar\'{e}
sphere representation. The Cartesian coordinates of this point are
the Stokes parameters $\{S_1, S_2, S_3\}$, where $S_0\equiv1$. $S_1$
represents the linear horizontal and vertical polarizations
$|h\rangle$ and $|v\rangle$, $S_2$ represents the linear diagonal
polarizations $|p\rangle=(|h\rangle+|v\rangle)/\sqrt{2}$, and
$|m\rangle=(-|h\rangle+|v\rangle)/\sqrt{2}$, and $S_3$ corresponds
to the circular polarizations
$|r\rangle=(|h\rangle+i|v\rangle)/\sqrt{2}$ and
$|l\rangle=(|h\rangle-i|v\rangle)/\sqrt{2}$). Points inside the
Poincar\'{e} sphere represent partially polarized states.

Consider an arbitrary quantum channel $\mathcal{E}$ that acts on a
single-qubit state $\hat{\rho}$. $\mathcal{E}$ is complete positive
and linear. It can be represented as the mapping of the surface of
the Poincar\'{e} sphere onto an ellipsoid which is contained within
the sphere. The operation of $\mathcal{E}$ can also be described
using the elements of the process matrix $\chi$
\begin{equation}\label{process_definition}
\mathcal{E}(\hat{\rho})=\sum_{m,n}\chi_{mn}\sigma_{m}\hat{\rho}\sigma_{n}^\dag,
\end{equation}
which is presented here in the basis of the identity matrix
$\sigma_0$, and the Pauli matrices $\{\sigma_1,\sigma_2,\sigma_3\}$.
$\chi$ is positive, Hermitian, and satisfies $\textrm{Tr}(\chi)=1$
(i.e., the channel is lossless). Channels that obey
$\mathcal{E}(\hat{I})=\hat{I}$ (mapping the maximally depolarized
state at the origin) belong to the unital channel class. In the
Poincar\'{e} sphere picture, unital channels are the case where the
mapped ellipsoid and the sphere are concentric. Dephasing channels
are a special case where the process can be written as
\begin{equation}\label{dephasing_channel}
\mathcal{E}(\hat{\rho})=(1-P)\hat{\rho}+P\sigma_3\hat{\rho}\sigma_{3}.
\end{equation}
They belong to the class of unital channels, and are characterized
by the probability $P$ to obtain a change in the phase of the
original state. A process matrix $\chi$ can describe a dephasing
process if it has two eigenvalues that equal zero. Denote the
highest eigenvalue of this $\chi$ by $\chi_0$, the probability $P$
of the corresponding dephasing process is:
\begin{equation}\label{dephasing_probability}
P=1-\chi_0.
\end{equation}
In order to implement a controllable dephasing channel that can vary
from no dephasing to a complete dephasing it is sufficient to show
that value of $P$ can have any value in the range of $\{0,0.5\}$.
This is because dephasing channels with higher $P$ values can be
represented as a combination of dephasing channels with $P$ that
lies in the above mentioned range, and another deterministic
bit-flip channel with a probability of $100\%$ that can easily be
built. It is clear from Eqs. (\ref{dephasing_channel}) and
(\ref{dephasing_probability}) that in the $\chi$ matrix
representation it is sufficient to show that a fully controllable
dephasing channel is obtained if the highest eigenvalue $\chi_0$ can
have any value in the range of $\{0.5,1\}$, and the two lowest
eigenvalues of $\chi$ remain zero for any $\chi_0$ settings.

A useful representation of unital single-qubit quantum processes is
the tetrahedron representation: The Choi-Jamio{\l}kowski isomorphism
between complete-positive linear maps and quantum states connects
between the process matrix of a single-qubit channel and the density
operator of a two-qubit state \cite{Jamiolkowski_1972}. Thus, we can
represent the 4-dimensional (4D) $\chi$ matrix of a single-qubit
unital channel via the representation of the corresponding class of
two-qubit states \cite{Horodecki_1996a}
\begin{equation}\label{Chi_decomposition}
\chi=\frac{1}{4}\left(I\otimes{I}+\underset{m,n=1}{\overset{3}{\sum}}D_{mn}\sigma_{m}\otimes\sigma_{n}\right).
\end{equation}
Here, $D$ is a $3\times 3$ real matrix, that epitomizes all the
channel parameters. Ignoring rotations, the process of the channel
can be geometrically represented by the coordinate vector
$\vec{D}=\{D_1,D_2,D_3\}$, which is composed of the eigenvalues of
$D$. The complete positivity of the process is equivalent to the
requirements that
\begin{equation}\label{Legal_processes}
|D_i\pm{D_j}|\leq|1\pm{D_k}|,
\end{equation}
where $i\neq{j}\neq{k}$ \cite{King_2001}. Equation
(\ref{Legal_processes}) dictates that all allowed $\vec{D}$ vectors,
when drawn in a cartesian 3D coordinate systems, span the volume of
a tetrahedron. The values of $\{D_1,D_2,D_3\}$ are related to the
eigenvalues of the corresponding $\chi$
($\chi_0,\chi_1,\chi_2,\chi_3$) by
\begin{equation}\label{Radii chi eigenvalues}
D_i=\chi_0+\chi_i-\chi_j-\chi_k\,,
\end{equation}
where $i\neq{j}\neq{k}\neq0$ \cite{King_2001}. It can be shown that
the eigenvalues of $D$ are equal to the lengths of the three primary
radii of the mapped ellipsoid in the Poincar\'{e} sphere
representation of the process. In the tetrahedron representation,
the identity process ($\mathcal{E}(\hat{\rho})=\hat{\rho}$)) is
represented by the $\{1,1,1\}$ vertex of the tetrahedron, and any
arbitrary dephasing channel with a corresponding probability $P$ is
represented by a point that lies on the tetrahedron edge of
$\vec{D}=\{1-2P,1-2P,1\}$.

\section{Experimental setup}

\begin{figure}[tbp]\noindent \begin{centering}
\includegraphics[width=100mm]{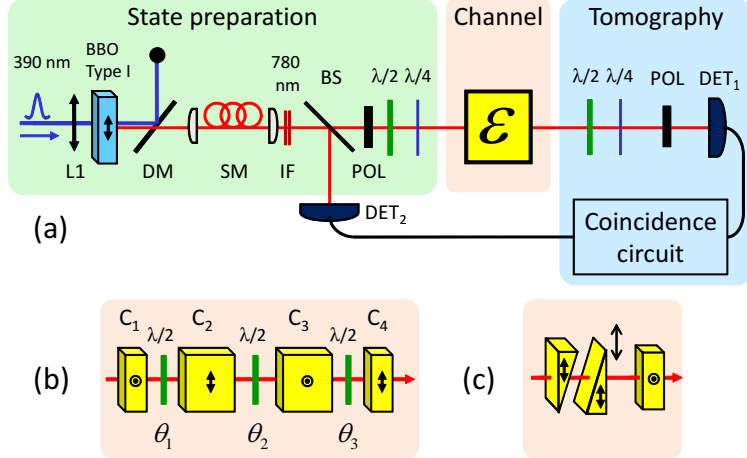}
\par\end{centering}
\caption{\label{Fig_QPT_Experimental_setup_scheme}(a) The
experimental setup for the characterization of single-qubit
channels. Photon pairs are generated in the BBO crystal, which is
located after a lens (L1). The down-converted signal is filtered by
passing through a dichroic mirror (DM), a single-mode fiber (SM) and
an interference bandpass filter (IF). The photon pairs are split
into two ports using a BS, where single-photon detectors (DET$_1$,
DET$_2$) are located at the end of each port. The investigated
channel ($\mathcal{E}$) is applied only on photons that emerge from
one of the ports. The initial polarization state of these photons is
determined using a polarizer (POL), a HWP ($\lambda/2$) and a QWP
($\lambda/4$), and their final polarization state is measured using
a HWP, a QWP and a polarizer that are placed after the channel. (b)
The four-crystal scheme, composed of four perpendicularly-oriented
calcite crystals and three HWPs. The thickness of the two outer
(inner) crystals is 1\,mm (2\,mm). (c) A Soleil-Babinet compensator,
composed of two translatable quartz wedges and another rectangular
fixed quartz crystal (see details in the appendix).}
\end{figure}

The experimental setup which was used to characterize the
implementation of the dephasing channel is presented in Fig.
\ref{Fig_QPT_Experimental_setup_scheme}(a). A pulsed 390\,nm pump
laser is focused onto a 1\,mm thick type-I
$\beta-\textrm{BaB}_{2}\textrm{O}_{4}$ (BBO) crystal. Photon pairs
are collinearly generated via the process of spontaneous parametric
down conversion. After the BBO crystal, the down converted signal is
separated from the 390\,nm pump beam by passing through a dichroic
mirror. Spatial and temporal filtering of the 780\,nm photon pairs
is achieved by sending the photon pairs through a single-mode fiber,
and a 5\,nm interference bandpass filter, respectively. Then, the
two photons are split probabilistically using a beam splitter (BS),
where one photon is sent directly to a detector, and the second
photon is directed to the investigated single-qubit channel. The
initial polarization state of the photon that is sent to the channel
is set using a polarizer, a half- and a quarter-wave plates (HWP and
QWP). After the channel, the output polarization state is measured
using a standard quantum state tomography (QST) procedure
\cite{James_2001}: the photon passes a QWP, a HWP and a polarizer
before being coupled to the detector.

We performed a complete QPT for every channel setting by
characterizing the channel effect on $|h\rangle$, $|v\rangle$,
$|p\rangle$, and $|r\rangle$ polarization states \cite{Chuang_1997}.
When only the counts of photons that pass through the dephasing
channel (counts from $\textrm{DET}_1$) are considered, the channel
effect on a classical single-photon state (i.e., a weak signal
photonic state that has the thermal statistics of the down
conversion process) is measured. When the detection of the photon
that experiences the dephasing noise is also conditioned in the
detection of another photon that does not pass through the channel
(i.e., coincidence counts from both $\textrm{DET}_1$ and
$\textrm{DET}_2$ are counted), the channel effect on quantum
single-photon states that have different temporal coherence
properties is investigated. The classical and the quantum
single-photon wave-packet states differ in their temporal
properties, as the quantum wave-packet has a longer coherence time
than classical wave-packet. In order to verify this difference, we
studied the effect of a dephasing scheme which is composed of a SBC
(see Fig. \ref{Fig_QPT_Experimental_setup_scheme}(c)) on the
classical and the quantum wave-packets \cite{Branning_2000}. The
dephasing results which clearly demonstrate the difference between
the two wave-packet types are presented in the appendix. The typical
count rate of single counts (photon hits in one detector
disregarding the other one) was $\sim20,000$\,Hz, and the rate of
the coincidence counts was $\sim1000$\,Hz. Single counts where
considered and analyzed after the subtraction of a stray light
signal noise (background counts were estimated to be in the order of
$2000$\,Hz). As for the coincidence counts, no subtraction was
required since the stray light signal noise was negligible. Errors
were calculated using a maximal likelihood procedure and Monte Carlo
simulations, assuming Poissonian noise for the photon counts
\cite{Kwiat_Tomo,Shaham_QPT}.

\section{Four-birefringent crystal scheme as a dephasing channel}

In order to apply decoherence to polarization qubits, one should
entangle the polarization degree of freedom (DOF) with extra DOFs
that are not going to be measured, effectively increasing the dimension of the state Hilbert space. Ignoring the extra DOFs, the measured density
matrix of the polarization DOF is obtained after tracing out these extra DOFs. Such a coupling between
the polarization and additional temporal DOFs of the wave-packet can be
achieved using a depolarizer that is composed of birefringent crystals and wave-plates in between them
\cite{Kwiat_Dephasing,Shaham_2011}. Consider a polarized wave-packet
of photons with a coherence time $\tau$ that passes through such a
depolarizer. A crystal with a length $L$ induces a temporal
delay $t=L\frac{\Delta{n}}{c}$ between the fast and the slow
polarization modes, where $\Delta{n}$ is the birefringent index
difference and $c$ is the speed of light. We require that all
crystals are sufficiently long such that the temporal delay between
different polarization modes of the output states is much larger than the
coherence time of the photons $\tau$:
\begin{equation}\label{temporal delay}
L\frac{\Delta{n}}{c}\gg\tau.
\end{equation}
After the passage through the depolarizer, the wave-packet occupies
a discrete number of temporal modes. These modes do not overlap in
between them, and every different temporal mode is fully polarized
by itself. The photon-detectors are insensitive to small temporal
differences between the different modes and do not record them. As a
result, the temporal DOFs of the photonic state are traced out and
decohered mixed states are detected. The mixture level of the
detected state depends only on the occupation of each temporal mode
and is a function of the input polarization state, and the relative
angle between the polarization state and the primary axes of the
crystal \cite{Shaham_2011}. It does not depend on the temporal shape
of the discrete modes which is inherited from the input temporal
shape and properties. Thus, as long as the relation of Eq.
(\ref{temporal delay}) is kept, the decoherence process that is
induced by such a depolarizing scheme is the same regardless of the
exact temporal shape of the photonic wave-packet. Turning the
wave-plates between the crystals affects the relative occupation of
each different temporal mode and controls the decoherence properties
that are to be measured.

We study a decohering scheme that is composed of four fixed calcite
crystals ($\textrm{C}_1,..,\textrm{C}_4$), and three tunable HWPs
($\theta_1,\theta_2,\theta_3$) in between them (see Fig.
\ref{Fig_QPT_Experimental_setup_scheme}(b)) \cite{Shaham_iso_depo}.
The length of $\textrm{C}_2$ and $\textrm{C}_3$ is 2\,mm, and the
length of $\textrm{C}_1$ and $\textrm{C}_4$ is 1\,mm. The fast axes
of $\textrm{C}_1$ and $\textrm{C}_3$, and the slow axes of
$\textrm{C}_2$ and $\textrm{C}_4$ are parallel, and define the zero
angle of the wave-plates. This scheme couples between the
polarization DOF and seven possible temporal modes: recall that a
1\,mm crystal induces a time delay of $\Delta{t}$ between the fast
and the slow polarization modes, the maximal time delay is obtained
between a polarization mode that travels through the fast axis of
every crystal in the configuration, and the polarization mode that
travels through the slow axis of every crystal, and equals
$(1+2+2+1)\Delta{t}=6\Delta{t}$. Polarization modes that travel
through fast axis of some crystals and the slow axis of the other
ones occupy all temporal modes which are the multiplications of
$\Delta{t}$ that lie between $t=0$ and $t=6\Delta{t}$, so that seven
different temporal modes are obtained (for a detailed mathematical
description of the different temporal modes see also
\cite{Shaham_2011}). Hence, a polarization qubit that passes through
this scheme resides within a 14-dimensional time-polarization
Hilbert space. For photons with a wavelength of 780\,nm, the
temporal delay between two successive modes is $\sim570$\,fs. The
coherence time $\tau$ of the down converted single-photon
wave-packets before entering the scheme is $\sim180\,\textrm{fs}$
(it is mainly governed by the 5\,nm interference bandpass filter).
Thus, the requirement of Eq. (\ref{temporal delay}) is fulfilled and
the seven different temporal modes can be regarded as discrete ones.

Tuning the HWPs to different angles, different processes are induced
by the scheme. The values of the different $D_i$ parameters (Eq.
(\ref{Radii chi eigenvalues})) that represent these processes as a
function of the HWP angles are
\begin{eqnarray}
D_1 &=& -\sin(4\theta_1)\sin(4\theta_3)\cos^2(2\theta_2)+\cos(4\theta_1)\cos(4\theta_2)\cos(4\theta_3), \label{eq_four_crystal_D1}\\
D_2 &=& \sin(4\theta_2)\sin(2\theta_1+2\theta_3)\cos(2\theta_1)\cos(2\theta_3)-\cos^2(2\theta_2)\cos^2(2\theta_1+2\theta_3)-\nonumber\label{eq_four_crystal_R2}\\
& &-\frac{1}{2}\sin(4\theta_1)\sin(4\theta_3)\sin^2(2\theta_2), \label{eq_four_crystal_D2}\\
D_3
&=&-\sin(4\theta_2)\sin(2\theta_1+2\theta_3)\cos(2\theta_1)\cos(2\theta_3)-\cos^2(2\theta_2)\cos^2(2\theta_1+2\theta_3)-\nonumber\label{eq_four_crystal_R3}\\
&
&-\frac{1}{2}\sin(4\theta_1)\sin(4\theta_3)\sin^2(2\theta_2).\label{eq_four_crystal_D3}
\end{eqnarray}
We emphasize that these values are calculated up to rotations that
flip the sign of two $D_i$ values. Cancelation of these rotations
can be achieved by adding more fixed wave-plates before the
crystals, or by tilting one of the crystals to change the induced
birefringent phase.

\begin{figure}[tbp]\noindent \begin{centering}
\includegraphics[width=90mm]{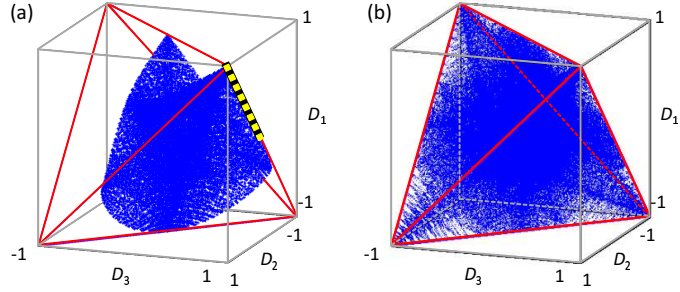}
\par\end{centering}
\caption{\label{Fig_Four_crystal_D_phase_diagraml} Numerical
investigation of the possible channels of the four-crystal scheme in
the tetrahedron representation. (a) All possible channels using
different orientations of the three HWPs of the four-crystal scheme.
(b) All possible channels using the different orientations of the
three HWPs of the four-crystal scheme, with the addition of all
allowed polarization rotations. These rotations can be implemented
using more wave-plates that are placed after the scheme. Points on
the dashed yellow line along the edge of the tetrahedron in (a)
correspond to a dephasing channel that preserves the length of
$D_3$.}
\end{figure}

Using Eqs. (\ref{eq_four_crystal_D1})- (\ref{eq_four_crystal_D3}),
we numerically investigated the spanned volume in the tetrahedron
representation that represents all possible processes that can be
emulated by this scheme. The spanned volume is shown in Fig.
\ref{Fig_Four_crystal_D_phase_diagraml}(a). The presented possible
process range can be extended with the addition of polarization
rotations that are described by two transformations: cyclic
permutations between the three $D_i$ values, and sign-flips of two
$D_i$ values. The extended process range is presented in Fig.
\ref{Fig_Four_crystal_D_phase_diagraml}(b). It can be seen from Fig.
\ref{Fig_Four_crystal_D_phase_diagraml}(b) that almost every
complete positive unital qubit map can be implemented using the
investigated four-crystal scheme.

Observing Fig. \ref{Fig_Four_crystal_D_phase_diagraml}(a), it can be
inferred that a dephasing channel can be implemented using this
scheme, since the spanned volume almost covers the edges of the
tetrahedron. However, a careful examination reveals that only an
approximation of the dephasing channel can be implemented. It can be
proved that there is no solution to Eqs. (\ref{eq_four_crystal_D1})-
(\ref{eq_four_crystal_D3}), that preserves the length of one of the
$D_i$ values equal to 1, while the absolute values of the two other
parameters have values in the range of (0,1). Thus, points on the
edges which correspond to a dephasing process are not attainable. We
examined the properties of the points that are in the vicinity of
the edges, and found that an approximation to the dephasing channel
is obtained for the following relation between the three HWPs angles
of the four-crystal scheme:
\begin{equation}\label{eq_four_crystal_dephasing_condition}
\theta_1=\theta_3=\frac{\theta_2}{2}.
\end{equation}

\begin{figure}[t]\noindent \begin{centering}
\includegraphics[width=100mm]{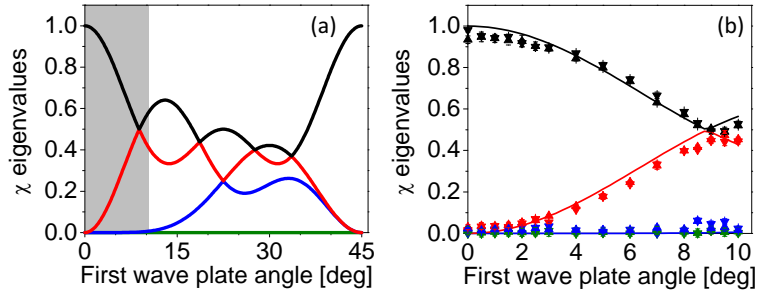}
\par\end{centering}
\caption{\label{Fig_dephasing_channel_four_crystal_results}
Four-crystal scheme as a dephasing channel. (a) Calculated
eigenvalues of the $\chi$ matrix as a function of the first HWP
angle. (b) Measured eigenvalues of the $\chi$ matrix as a function
of the first HWP angle for the dephasing range
($0\leq\theta_1\leq10^\circ$, gray area of (a)). Eigenvalues that
were reconstructed from counts of a single detector (coincidence
measurements) are presented as upright triangles (upside-down
triangles). Solid lines are the theoretical prediction.}
\end{figure}

Maintaining this relation, we calculated the corresponding process
matrices $\chi$ for $0\leq\theta_1\leq45^\circ$. A plot of the
eigenvalues of $\chi$ as a function of $\theta_1$ is presented in
Fig. \ref{Fig_dephasing_channel_four_crystal_results}(a). One of the
eigenvalues remains zero for every angle setting, and a second one
is approximately zero for small $\theta_1$ angles (see the gray area
in Fig. \ref{Fig_dephasing_channel_four_crystal_results}(a)). The
two other eigenvalues participate in the process for every angle,
and intersect when $\theta_1\simeq9^\circ$. Thus, an approximation
to a dephasing channel is obtained when $\theta_1$ is in the range
of $0\leq\theta_1\leq9^\circ$, and the other two HWP angles satisfy
Eq. (\ref{eq_four_crystal_dephasing_condition}).

We denote the highest eigenvalue of $\chi$ by $\chi_m$. Using Eqs.
(\ref{dephasing_channel}), (\ref{dephasing_probability}),
(\ref{Radii chi eigenvalues}), (\ref{eq_four_crystal_D1})-
(\ref{eq_four_crystal_D3}) and
(\ref{eq_four_crystal_dephasing_condition}), the dephasing
probability $P$ that is induced by the approximated dephasing
channel can be written as a function of $\theta_1$:
\begin{equation}\label{eq_four_crystal_dephasing_probability}
P=1-\chi_m=\frac{-3\cos^4(4\theta_1)+2\cos^2(4\theta_1)+1}{2}.
\end{equation}
In a similar manner to the definition of the fidelity between two
quantum states \cite{Jozsa_1995_fidelity}, the fidelity $F$ between
two different processes can be defined as
\begin{equation}\label{eq_process_fidelity_definition}
F(\hat{\chi}_{1},\hat{\chi}_{2})=\left(\mathrm{\textrm{Tr}}\left(\sqrt{\sqrt{\hat{\chi}_{1}}\hat{\chi}_{2}\sqrt{\hat{\chi}_{1}}}\right)\right)^{2}.
\end{equation}
Comparing between a theoretical ideal controlled dephasing channel
and the theoretical prediction of the approximated dephasing
processes in the range $0\leq\theta_1\leq9^\circ$ that have the same
corresponding $P$, the minimal value of the calculated process
fidelity is $99.6\%$. Actually, such a high value is better than the
typical accuracy that is achieved in experimental realizations of
such channels. It is important to mention that the induced processes
are not accompanied by nontrivial rotations: defining the $S_1$
polarization directions as the directions of primary axes of the
crystals, the channel maintains the $S_2$ value of the input light,
and reduces the absolute values of $S_1$ and $S_3$. It also flips
the signs of the $S_2$ and $S_3$ parameters - a sign flip that can
be compensated for by adding a fixed HWP in a zero angle before or
after the channel.

Implementing the four-crystal scheme (Fig.
\ref{Fig_QPT_Experimental_setup_scheme}(b)), we measured the
processes that approximate the dephasing channel using the setup of
Fig. \ref{Fig_QPT_Experimental_setup_scheme}(a). Figure
\ref{Fig_dephasing_channel_four_crystal_results}(b) presents the
measured eigenvalues of the $\chi$ matrices that were reconstructed
from the data of classical single-photon wave-packets, as well as
those who were reconstructed from the data of quantum wave-packets
(see details in the experimental setup section). The eigenvalues are
presented as a function of $\theta_1$ along with the theoretical
prediction. The measured results are in good agreement with theory,
and the control over the dephasing level from no dephasing
($\theta_1=0$) to a complete dephasing ($\theta_1\simeq9^\circ$) is
achieved. As expected, no significant differences between the
dephasing processes of wave-packets with different temporal
properties are observed. Neglecting channel rotations, the average
fidelity between the measured processes and the theoretical
dephasing processes (which were calculated using Eq.
(\ref{eq_four_crystal_dephasing_probability}) for the corresponding
$\theta_1$ angle) is $97\pm2\%$, both for the processes which were
applied to classical and to quantum single-photon wave-packets. For
light of a broader spectral bandwidth we expect to obtain even
better results: such a wave-packet has a shorter coherence time and
the requirement of Eq. (\ref{temporal delay}) is easily fulfilled
since the overlap between different temporal modes after the passage
through the crystal is more negligible. Thus, Eqs.
(\ref{eq_four_crystal_D1})-(\ref{eq_four_crystal_D3}) more
faithfully represent the decoherence of the polarization state, and
a similar experiment will show a higher fidelity to the theoretical
calculation when compared to the presented classical and quantum
wave-packet cases.

\section{Conclusions and discussion}
To summarize, we investigated a controllable photonic dephasing
channel of a single spatial mode that is composed of four
birefringent crystals and wave-plates. This configuration can
emulate almost every arbitrary unital channel. The operation of the
channel as a dephasing channel was experimentally demonstrated, and
was characterized using single-photon wave-packets with different
coherence times. There is a high fidelity between the operation of
the explored scheme and that of an ideal dephasing environment,
where the dephasing rate is in agreement with the theoretical
prediction. As expected, the noise probability of the four crystal
scheme is known in advance and is not affected by changes in the
temporal envelope of the photonic wave-packet. The presented scheme
is not limited to single-photon states. It can be applied on a
high-intensity classical light of a short coherence time (or a
broader bandwidth), and serve as a depolarizer that emulates almost
any unital process. In the case of a wave-packet that has a longer
coherence time, the birefringent crystals can be replaced by
polarization maintaining birefringent fibers which act as longer
birefingent crystals.

\section*{Appendix: SBC as a dephasing channel for classical and quantum single-photon wave-packets}

In order to show that the classical and quantum single-photon
wave-packets differ in their fine temporal properties and that this
difference affects the dephasing level that these wave-packets
experience when transmitted through a common dephasing channel, we
transmitted the same investigated states through a SBC dephasing
scheme \cite{Branning_2000} (see Fig.
\ref{Fig_QPT_Experimental_setup_scheme}(c) in the main text). The
SBC dephasing scheme is composed of two birefringent crystal prisms
with parallel optical axes, which may be followed by another
birefringent rectangular crystal. The passage through the crystal
prisms results in a temporal delay of $t=L\frac{\Delta{n}}{c}$
between the two polarization modes (here $L$ is the \emph{total}
optical path inside the birefringent medium). When one of the prisms
is transversely translated, $L$ is changed, and a control over the
time delay $t$ is attained. When $t$ is comparable or larger than
the initial wave-packet coherence time $\tau$, the temporal overlap
between the two polarization modes is reduced and the phase
uncertainty between the two polarizations increases (i.e., the
photons experience a dephasing process). If a rectangular crystal is
added to the beam path (see Fig.
\ref{Fig_QPT_Experimental_setup_scheme}(c)), such that its optical
axes are perpendicularly oriented with respect to the two prisms,
the polarization modes can be temporally overlapped again and a
dephasing process with a small or a zero probability is emulated.

The SBC dephasing channel was implemented using two wedge quartz
crystals with a wedge angle of $15^{\circ}$, and another 9\,mm
rectangular quartz crystal. The chosen wedge angles of the crystals
enable a time delay up to $t\sim380\,\textrm{fs}$. Such a time delay
is larger by almost 2 orders of magnitude than the maximal time
delay that a typical Soleil-Babinet compensator can induce. As was
mentioned before, the coherence time $\tau$ of the down converted
photons is $\sim180\,\textrm{fs}$. As a result, we could
continuously switch between different dephasing levels. For every
translation setting, $t$ was calculated using the optical path
inside the birefringent quartz medium. We define $t$ to be in the
range of $0<t<\sim380\,\textrm{fs}$ when the primary axes of the
rectangular crystal are perpendicular to those of the wedge
crystals. The range of $t$ is extended up to 900\,fs using different
settings of the third rectangular crystal: time range of
$\sim270\,\textrm{fs}<t<\sim650\,\textrm{fs}$ is obtained when the
rectangular crystal is omitted, and time range of
$\sim530\,\textrm{fs}<t<\sim920\,\textrm{fs}$ is attained when the
rectangular crystal primary axes are parallel to those of the wedge
crystals.

\begin{figure}[tp]\noindent \begin{centering}
\includegraphics[width=100mm]{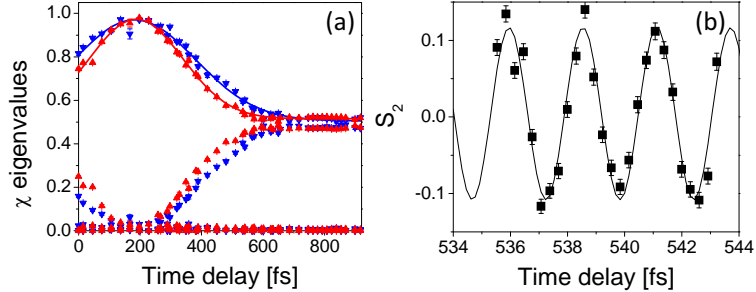}
\par\end{centering}
\caption{\label{Fig_Dephaser_chi_eigenvalues_and_s2_oscillations}
(a) The eigenvalues of the $\chi$ matrix as a function of the time
delay $t$ between the two polarization modes. Eigenvalues of process
matrices that were reconstructed from counts of single detector
(coincidence measurements) are presented as red upright triangles
(blue upside-down triangles). Solid lines are Gaussian fits to lead
the eye. (b) The oscillations of the Stokes parameter $S_2$ as a
function of the time delay between the two polarization modes. Solid
line represents a sinusoidal fit.}
\end{figure}

Performing a standard QPT procedure on the channel, we reconstructed
the process matrix $\chi$ for different translation settings. The
eigenvalues of the reconstructed $\chi$ matrices are presented in
Fig. \ref{Fig_Dephaser_chi_eigenvalues_and_s2_oscillations}(a) as a
function of the induced temporal delay $t$. As in the four crystal
case, the processes were separately reconstructed from data of
classical and quantum wave-packet states. It can be seen that both
states experience a dephasing operation since for every $t$, two
$\chi$ eigenvalues remain close to zero. When
$t\simeq185\,\textrm{fs}$, the operation of the channel is very
close to having no effect. The corresponding measured processes from
both data sets for this temporal setting have fidelities higher than
$95\pm2\%$ to the no-decohering process (disregarding polarization
rotations). When compared to the quantum single-photon wave-packet,
it is clear from Fig.
\ref{Fig_Dephaser_chi_eigenvalues_and_s2_oscillations}(a) that the
classical single-photon wave-packet experiences a faster dephasing
process. This demonstrates the difference between the two
wave-packet states, and shows the dependence of the dephasing
probability of the SBC scheme in the temporal shape of the incoming
photonic state, unlike the four crystal scheme (see Fig.
\ref{Fig_dephasing_channel_four_crystal_results}(b) in the main
text).

It is worth mentioning that although the wedge angle of the channel
is larger with respect to that of a typical Soleil-Babinet
compensator, the SBC dephasing scheme realization can also serve as
a solid and stable polarization interferometer. Figure
\ref{Fig_Dephaser_chi_eigenvalues_and_s2_oscillations}(b) presents
the oscillations of the $S_2$ Stokes parameter as a function of the
temporal delay. The oscillations were measured for the quantum
single-photon wave-packet state, when a heavy dephasing process is
applied to the photons ($P\sim0.45$). Analyzing the presented
oscillations, the calculated wavelength of the photons is
$775\pm15$\,nm, in agreement with the set value of 780\,nm.

\section*{Funding}
We thank the Israel Science Foundation for supporting this work
under Grants No. 546/10 and 793/13. We also thank the Israeli
Ministry of Science and Technology for financial support through the
Eshkol fellowship for A.S..


\end{document}